\documentclass[prl,twocolumn,letter, showpacs, superscriptaddress]{revtex4}
\usepackage{amssymb}
\usepackage{amsmath}
\usepackage{bbm}
\usepackage{graphicx}

\begin{document}

\title{A continuous-time solver for quantum impurity models}
\author{Philipp Werner}
\affiliation{Department of Physics, Columbia University, 538 West, 120th Street, New York, NY 10027, USA}
\author{Armin Comanac}
\affiliation{Department of Physics, Columbia University, 538 West, 120th Street, New York, NY 10027, USA}
\author{Luca de' Medici}
\affiliation{Department of Physics, Columbia University, 538 West, 120th Street, New York, NY 10027, USA}
\affiliation{Centre de Physique Th\'eorique, Ecole Polytechnique, 91128 Palaiseau Cedex, France}
\author{Matthias Troyer}
\affiliation{Institut f{\"u}r theoretische Physik, ETH H{\"o}nggerberg, CH-8093 Z{\"u}rich, Switzerland}
\author{Andrew J. Millis}
\affiliation{Department of Physics, Columbia University, 538 West, 120th Street, New York, NY 10027, USA}

\date{July 12, 2006}

\hyphenation{}

\begin{abstract}
We present a new continuous time solver for  quantum impurity models such as those relevant 
to dynamical mean field theory. It is based on a stochastic sampling of a 
perturbation expansion in the impurity-bath hybridization parameter. 
%For the one-orbital Hubbard model, no sign problem is found. 
Comparisons to quantum Monte Carlo and exact diagonalization calculations confirm the 
accuracy of the new approach, which allows very efficient simulations 
even at low temperatures and for strong interactions. As examples of the power of the method
we present results for the temperature dependence of the kinetic energy and the free
energy, enabling an accurate location of the temperature-driven metal-insulator transition.
\end{abstract}


\maketitle

Numerical computation of dynamical properties of strongly correlated fermion systems is one of the 
fundamental challenges in condensed matter physics. The exponential growth of the Hilbert space 
renders exact diagonalizations impossible except for very small systems \cite{exactdiag}. When 
applied directly to lattice models, auxiliary field Monte Carlo methods encounter 
severe difficulties with the fermionic sign problem \cite{auxiliary_problem}. The density matrix
reormalization group \cite{White92} is powerful for extracting ground state properties 
of one dimensional systems, but extensions to higher dimensions and dynamics have 
proven difficult. Over the last decade, a promising new ``dynamical mean field" (DMFT) 
approach has been developed \cite{Mueller-Hartmann89, Metzner89, Georges92}. 
As shown by Georges, Kotliar and other workers \cite{Georges92,Georges96,Maier05}, if the 
momentum dependence of the self-energy is neglected, 
$\Sigma(p,\omega)\rightarrow\Sigma(\omega)$, then the solution of the lattice model may be 
obtained from the solution of a quantum impurity model plus a self-consistency condition. 

A quantum impurity model represents an atom or molecule embedded in a host medium; in addition to their relevance to DMFT calculations these models are of intrinsic interest
and important to nano-science as representations of single-molecule conductors.
A quantum impurity model consists of a set of levels $a$ (with creation operators $\psi_a^\dagger$), populated by electrons which interact via general four-fermion interactions $U^{abcd}$ (for example density-density or spin exchange terms). The levels $a$ are hybridized to  
electronic continua (``bath orbitals") representing the degrees of freedom of the host material. These bath degrees of freedom may be integrated out, and the impurity model 
specified by a partition function $Z=\text{Tr} T_\tau e^{-S}$ with effective action $S=S_F+S_\text{loc}$, where
\begin{eqnarray}
S_F &=& -\int_0^\beta d\tau d\tau' \psi_a^{\phantom\dagger}(\tau) F_a(\tau-\tau')\psi_a^\dagger(\tau'), \label{S_F}\\
S_\text{loc} &=& -\int_0^\beta d\tau (\epsilon^{ab}\psi_a^\dagger\psi_b^{\phantom\dagger}-U^{abcd}\psi^\dagger_a\psi^\dagger_b\psi^{\phantom\dagger}_c\psi^{\phantom\dagger}_d). \label{S_loc}
\end{eqnarray}
%
%The indices of the spinors $\psi$ label orbital and spin. 
%Note that we have chosen the creation operator basis rather than the more conventional coherent state basis and an unconventional operator ordering in Eq.~(\ref{S_F}). 
%Our hybridization 
The function $F$ 
is determined by the hybridization and bath density of states. It
describes transitions from the impurity into the bath and back and is related to the mean field function $G_0^{-1}$ \cite{Georges96} by $G_0^{-1}(i\omega)=i\omega+\mu-F(-i\omega)$.
A solution of the quantum impurity model amounts to evaluating the Green functions $G(\tau)=-\langle T_\tau\psi(\tau)\psi^\dagger(0)\rangle$ for given $F$, $\epsilon$ and $U$. The dynamical mean field procedure involves a self-consistent determination of the hybridization function $F$, in which the computation of $G$ is time critical \cite{Georges96}. 

Perhaps the most commonly used technique is
the Hirsch-Fye  method \cite{Hirsch86}, in which a (discrete) Hubbard-Stratonovich 
transformation is used to decouple  the interaction part, leading 
to determinants which give the weights associated with the 
configurations of the auxiliary fields, which are then sampled by a Monte Carlo procedure. 
The method requires a discretization of the imaginary time interval $[0, \beta]$ into $N$ equal slices
and the evaluation of determinants of  $N\times N$  matrices.
At low temperatures and strong correlations, the Green functions have a highly non-uniform 
time dependence, dropping very steeply near the edge points $\tau=0,\beta$, and varying more slowly 
in between. The rapid drop means that a very large $N$ ($\sim 5U\beta$) is required, limiting  the 
utility of the algorithm for physically relevant temperatures. Furthermore, useful decouplings of the general
interactions of multiorbital models are lacking. Another approach is an exact diagonalization method \cite{Caffarel94,Toschi05} in which the continuum of energies in the impurity model is replaced by a 
discrete set of levels. This method suffers from  systematic errors associated with the 
discrete level spacing, and becomes impractical  in the multiorbital case because 
of the rapid growth of the Hilbert
space.

In this paper we present a new, continuous time approach which resolves many of these issues.  The idea,  
introduced in the context of bosonic field theories by Prokof'ev {\it et al.}~\cite{Svistunov96}, 
is to define a formal perturbation expansion for the partition function $Z$. Certain collections of terms in this series
are then sampled by a Monte Carlo procedure. In important prior work, Rubtsov and 
co-workers \cite{Rubtsov05} performed such a stochastic evaluation
by expanding in the interaction term. In this paper we show 
how to formulate a determinental Monte Carlo algorithm by
expanding in the hybridization term $S_F$
while  treating the interactions exactly.
This strong-coupling approach becomes particularly powerful at the strong interactions characteristic of systems of present day interest (such as high-$T_c$ cuprates),
because the perturbation order actually decreases with increasing $U$. 
Our method allows access to low enough temperatures that  both ground state properties
and the leading temperature dependencies may be determined, even at very strong couplings, 
providing new information unavailable by other methods.

\begin{figure}[t]
\centering
\includegraphics [angle=0, width= 7.5cm] {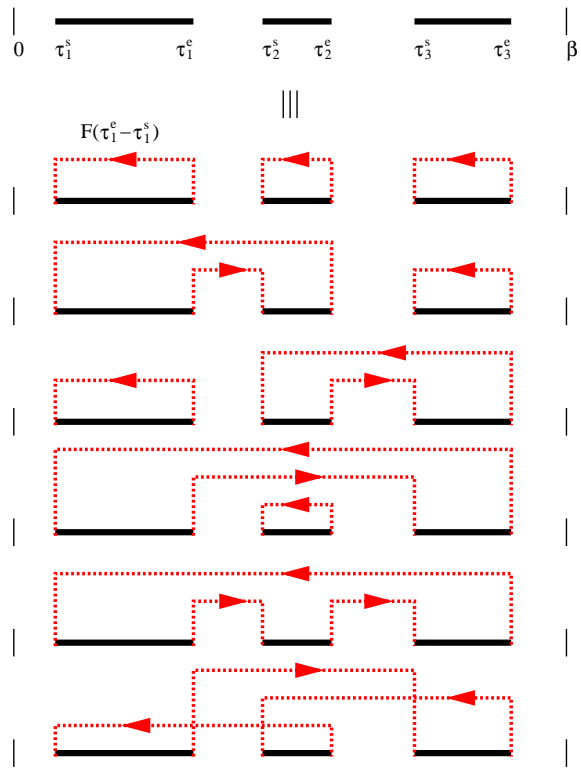}
\caption{Illustration of $Z^F_3(\tau_1^s, \tau_1^e; \tau_2^s, \tau_2^e; \tau_3^s, \tau_3^e)$ in terms of 
diagrams. A full line corresponds to an occupied fermion state. Each end point $\tau_m^e$ of a 
segment is connected to a starting point $\tau_n^s$ by a dashed line representing 
$F(\tau_m^e-\tau_n^s)$. %Diagrams with an odd number of crossing dashed lines have negative weight.
The combined weight of the diagrams can be expressed as a determinant.}
\label{diagrams}
\end{figure}

We illustrate the procedure first for noninteracting spinless fermions, and then proceed to the one-orbital Hubbard model, reserving a general discussion for a future publication \cite{Werner06b}. Because the spinless fermion model is diagonal in the occupation number basis, we have for $Z^{F} \equiv Z(\mu=0,U=0, \sigma)$ 
\begin{align}
&Z^{F}=2+\sum_{k=1}^\infty \int_{0}^\beta d\tau_1^s \int_{\tau_1^s}^{\beta}d\tau_1^e \int_{\tau_1^e}^{\beta} d\tau_2^s \int_{\tau_2^s}^{\beta}d\tau_2^e \ldots\nonumber\\
&\ldots \int_{\tau_{k-1}^e}^{\beta} d\tau_k^s \int_{\tau_k^s}^{\circ\tau_1^s}d\tau_k^e Z^F_k(\tau_1^s,\tau_1^e;\tau_2^s,\tau_2^e;\ldots;\tau_k^s,\tau_k^e),\hspace{0mm}
\label{series}
\end{align}
with $Z^F_k(\tau_1^s,\tau_1^e;\tau_2^s,\tau_2^e;\ldots;\tau_k^s,\tau_k^e)$ denoting the 
collection of $k!$ diagrams containing $k$ segments of occupied fermion states, with starting 
point $\tau_i^s$ and end point $\tau_i^e$, $i=1,\ldots, k$. We view the interval $[0,\beta]$ as a 
circle and take care of the trace
over occupation number states by including diagrams in which the last fermion line winds around the 
circle. This is denoted by an upper integral bound $\circ\tau$. An illustration of these diagrams 
for $k=3$ is shown in Fig.~\ref{diagrams}. Each end point $\tau_m^e$ is connected to a starting 
point $\tau_n^s$ by a dashed line representing $F(\tau_m^e-\tau_n^s)$. The collection of 
diagrams evaluates to 
\begin{align}
&Z^F_k(\tau_1^s,\tau_1^e;\tau_2^s,\tau_2^e;\ldots;\tau_k^s,\tau_k^e) = \det M^{-1} \delta_{\tau_1^s}^{\tau_k^e}, \\ 
&(M^{-1})_{m,n} = F(\tau_m^e-\tau_n^s), \label{M_inv}
\end{align} 
with $\delta_{\tau_1^s}^{\tau_k^e}=-1$ if $\tau_k^e<\tau_1^s$ 
(the last segment winds around the circle
) and $+1$ otherwise. One also has to sample the ``full line" and the ``empty line" states, 
corresponding to 1 or 0 particle in the whole interval $0\le \tau < \beta$. 
We use three types of Monte Carlo updates:
(i) insertion and removal of a segment,
(ii) insertion and removal of an anti-segment, and
(iii) shift of the segment end point.
The first two  are required for ergodicity and the third enhances the 
sampling efficiency.
Combining the $k!$ diagrams into a determinant is essential. %Diagrams with an odd number of 
%crossing $F$-lines have negative weight (the simplest example is the last diagram shown in Fig.~\ref{diagrams}),
Some diagrams have negative weight and a ``worm"-type algorithm which sampled individual diagrams would run into a bad sign problem. Yet we found in all of our computations that the determinant remains positive. %%

Suppose we start with a collection of segments
$s_k = (\tau_1^s,\tau_1^e;\tau_2^s,\tau_2^e;\ldots;\tau_k^s,\tau_k^e)$ and 
want to insert a new segment $\tilde s$ at a randomly chosen $\tilde\tau^s$. 
If $\tilde \tau^s$ happens to lie on a segment of $s_k$, the move to the new configuration 
$s_{k+1}=s_k+\tilde s$ is rejected, otherwise it has to satisfy a detailed balance condition. 
If the length $\tilde l$ of the new segment is chosen randomly in the interval 
$[0,l_\text{max}]$, with $l_\text{max}$ determined by $\tilde \tau^s$ and the 
collection $s_k$ of existing segments, and we propose to remove this new 
segment in the new configuration $s_{k+1}$ with probability $1/(k+1)$, the condition is
\begin{align}
\frac{p_\text{ins}(s_\text{new})}{p_\text{rem}(s_\text{new})} =& 
\frac{Z^F_{k+1}(s_{k+1})}{Z^F_k(s_k)}\frac{\beta l_\text{max}}{k+1}e^{\tilde l\mu}.
\label{update_segment}
\end{align}
The result for anti-segments is similar and for shift updates from a 
configuration $s$ to a configuration $\tilde s$ (changing the length of some 
segment from $l$ to a randomly chosen $\tilde l\in[0,l_\text{max}]$) one obtains
\begin{align}
\frac{p(s\rightarrow \tilde s)}{p(\tilde s \rightarrow s)} =& 
\frac{Z^F_{k}(\tilde s)}{Z^F_k(s)}e^{(\tilde l - l)\mu}.
\label{update_shift}
\end{align}
\begin{figure}[t]
\centering
\includegraphics [angle=-90, width=8.5cm] {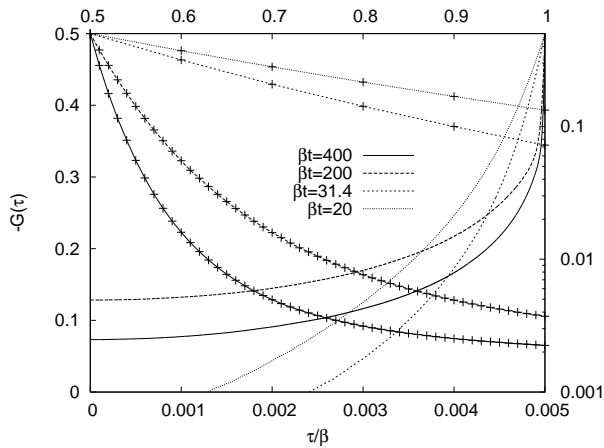}
\caption{Green functions for $n=1$, $U/t=3.5\sqrt{2}$, $\beta t=400, 200, 31.4$ and $20$. 
Lines without symbols (upper and right axes) show $G(\tau)$ on a semi-log scale over the 
wide time interval $[\beta/2,\beta]$ revealing marked differences between metallic ($\beta t=200,400$) 
and insulating ($\beta t=20,31.4$) solutions. Lines with symbols (lower and left axes) show the same 
data on a linear scale in the very narrow $\tau$ range $[0, \beta/2000]$, revealing the accurate 
representation of the rapid drop of $G(\tau)$. 
}
\label{drop}
\end{figure}
A procedure analogous to that in Ref.~\cite{Rubtsov05} is used to calculate
the determinant ratios and the new 
enlarged (reduced) matrices in a time $O(k^2)$.
We store and manipulate $M$,  the inverse of Eq.~(\ref{M_inv}), because $M$ allows easy access to
the determinant ratios in Eqs.~(\ref{update_segment}) and (\ref{update_shift}) and
is required for measuring the Green function, since
\begin{align}
G(\tau) &= \Big\langle\frac{1}{\beta}\sum_{i=1}^k\sum_{j=1}^k M_{j,i}\Delta(\tau, \tau_i^e-\tau_j^s)\Big\rangle,\\
\Delta(\tau, \tau') &= \left\{ \begin{array}{ll} \delta(\tau-\tau') & \tau'>0 \\
                                                             -\delta(\tau-\tau'-\beta) & \tau'<0 
                                     \end{array} \right..\label{Delta}\hspace{5mm}
\end{align}
The end points $G(0)$ and $G(\beta)$ can be measured accurately from the average total length of the segments. 

In the form given here, the algorithm generalizes straightforwardly to any model with interaction terms which are diagonal
in an occupation number basis (for models with exchange, see Ref.~\cite{Werner06b}).
One simply  introduces one collection of segments for
each spin/orbital state, and the weight of a configuration now also depends on the segment 
overlap. For example, in the one-orbital
Hubbard model with on-site interaction $U$, there is one collection of segments for
spin up and one for spin down, while in Eqs.~(\ref{update_segment}) 
and (\ref{update_shift}) one has to add a factor $\exp(-\delta_\text{ov}U)$ on 
the right hand side, where $\delta_\text{ov}$ denotes the change in overlap 
between up and down segments. 

We have used the new method to study the paramagnetic phase of the Hubbard model with semicircular
density of states of bandwidth $4t$, for interactions of the order of the Mott critical value $U_{c2}$
and temperatures as low as $\beta t=400$. For this model the 
self-consistency condition reduces to
$F(\tau) = t^2 G(-\tau)$.
Simulations for temperatures down to 
$\beta t\approx 50$ can be run on a laptop. For calculations at $\beta t = 400$, we 
typically used 10 CPU hours for each iteration in order to accurately resolve the short- 
and long-time behavior. 

Figure~\ref{drop} shows the impurity model Green function for 
$U/t=3.5\sqrt{2}$, $\beta t=20$, 31.4, 200 and $400$ and $n=1$ (half filling).
The lower two temperatures are out of reach of the Hirsch-Fye algorithm. 
We collected the data on a grid of $10^4$ points for $\beta t=200, 400$ and $10^3$ points for 
$\beta t=20, 31.4$. The lines with symbols
show that the method accurately captures the steep short-time drop of $G$; the lines without
symbols demonstrate clearly the difference in long-time behavior between the insulating
(high-$T$) and metallic (low-$T$) solutions.

\begin{figure}[t]
\centering
\includegraphics [angle=-90, width= 8.5cm] {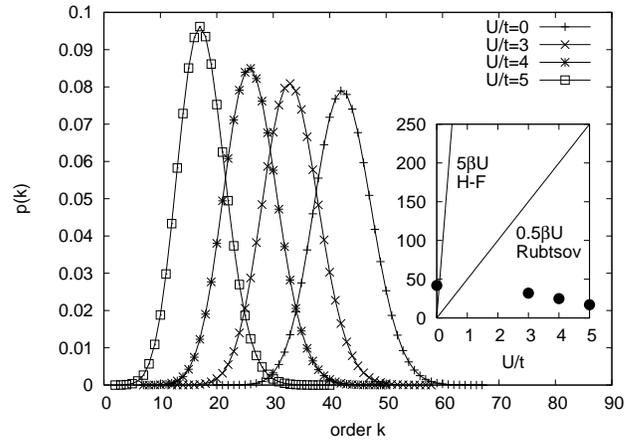}
\caption{Probability $p(k)$ for a configuration with $k$ segments plotted for different interaction strengths $U/t$ for $\beta t=100$ and half-filling. The peak position shifts to lower 
values of $k$ as $U/t$ is increased. The inset compares the scaling of the matrix size with $U$ to Hirsch-Fye ($\approx 5\beta U$) and the method of Ref.~\cite{Rubtsov05} ($\approx 0.5\beta U$).
}
\label{order}
\end{figure}

Despite the almost perfect resolution, the typical size, $k$, of the matices, $M$,
which are generated during the simulation remains reasonable even at low temperatures. 
This property explains the superior performance of the strong-coupling expansion method. 
Figure~\ref{order} 
shows the probability distribution $p(k)$ for $\beta t =100$ and different values of the interaction 
strength. While the peak value of the distribution is proportional to $\beta$, it shifts 
to \textit{lower} order as the interaction strength is increased, in contrast to Hirsch-Fye or 
the method of Ref.~\cite{Rubtsov05}, where the matrix size scales approximately as 
$5\beta U$ and $0.5\beta U$, respectively. The inset of Fig.~\ref{order} shows that the linear size of the matrix in our method can easily be a factor 100 smaller than in a Hirsch-Fye calculation or a factor $10$ smaller than in the weak-coupling approach of Ref.~\cite{Rubtsov05}.
The cubic scaling of the computational effort with matrix
size implies a dramatically improved efficiency at couplings of the order of the Mott critical value, making
low $T$ behavior accessible.
\begin{figure}[ht]
\centering
\includegraphics [angle=-90, width= 8.5cm] {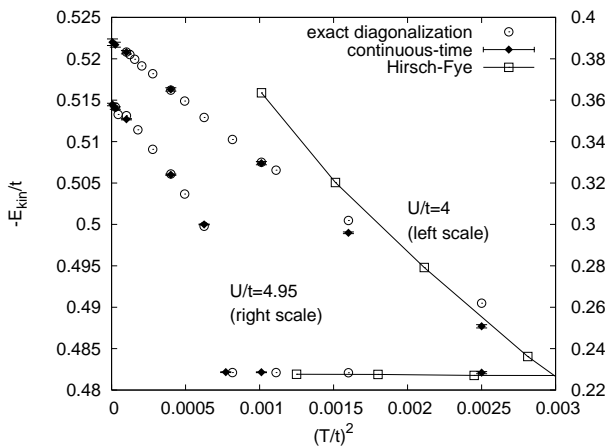}
\caption{Kinetic energy obtained using the indicated impurity solvers plotted as a 
function of temperature for $U/t=4$ and $U/t=3.5 \sqrt{2}\approx 4.95$. 
The former value corresponds to a metallic phase, while for $U/t=3.5 \sqrt{2}$, 
the system undergoes a metal-insulator transition.}
\label{kinetic}
\end{figure}

To verify the accuracy of the method we  show in Fig.~\ref{kinetic} the kinetic energy 
$K=2t^2\int_0^\beta d\tau G(\tau)G(-\tau)$ obtained via the new approach, the exact diagonalization method
\cite{Capone06}, and
the Hirsch-Fye method. The results are plotted against $T^2$ to show that the theoretically
expected Fermi liquid result can be obtained even when the characteristic energy scales are very low.
The new  algorithm displays the metal-insulator transition clearly,
agrees perfectly with  exact diagonalization results,
and is consistent with the Hirsch-Fye data at elevated temperatures. 

\begin{figure}[ht]
\centering
\includegraphics [angle=-90, width= 8.5cm] {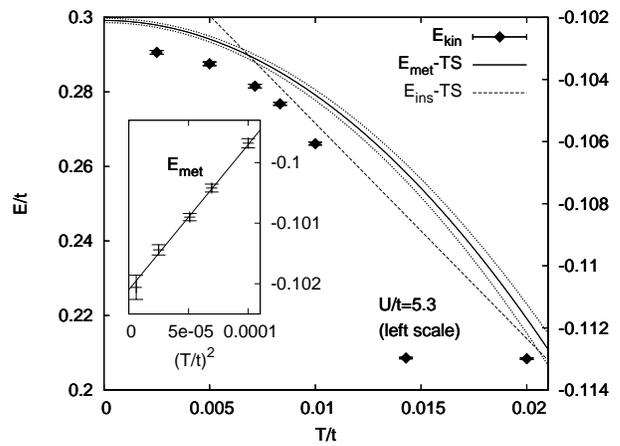}
\caption{Points: Kinetic energy as a function of temperature for $U/t=5.3$. Solid line with error estimates as dotted lines: free energy of the metallic state. Dashed line: free energy of the insulating state. Inset: 
quadratic $T$-dependence of the total energy.
}
\label{free}
\end{figure}

Figure~\ref{kinetic} shows a  discontinuity arising from the instability
of the metallic solution as the temperature is increased above a certain value ($T/t\approx 0.026$ at $U/t=4.95$).
The instability implies the existence of a first order metal-insulator transition, but 
strictly speaking it is a spinodal point.  
To further demonstrate the power of the method we locate the true phase boundary by
constructing the free energy of the metallic state. We compute the total internal energy 
per site $E(T)=K(T)+U\langle n_\uparrow n_\downarrow \rangle$, from
which we obtain the specific heat $C(T)=dE(T)/dT$ and hence 
the entropy  $S(T)=\int_0^TdT' (C(T')/T')$. We compare this to the free energy 
of the paramagnetic insulating state for which $E(T)$ has negligible
$T$-dependence because of the gap, but the entropic contribution is $-T\ln 2$. Results 
are illustrated in Fig.~\ref{free} for $U/t=5.3$, where the interaction is 
so large that exact diagonalization data in the relevant $T$ regime are not  available. The jump in the kinetic energy is clearly seen to be a spinodal effect, occuring at a higher temperature ($T \approx 0.011t$) than the
$T_c = 0.00685(4)t$ at which the free energies cross. The latent heat of this metal-insulator transition is $T_c(S_\text{met}(T_c)-S_\text{ins})=0.00250(5)t$. 

In conclusion, we have developed a continuous-time impurity solver, based on a stochastic evaluation
of an expansion in the hybridization. 
The new approach, which (even away from half-filling) does not appear to suffer from a sign problem, is  much more efficient at intermediate to strong couplings than other available methods.
It allows to simulate the 
Hubbard model at previously inaccessible temperatures, permitting direct access to the asymptotic
low energy physics, for example allowing the computation of the latent heat across the metal-insulator transition. The algorithm opens up for systematic study the low temperature properties of strongly interacting quantum models relevant to transition metal oxides, actinides and other correlated electron materials. 

PW, AC and AJM acknowledge support from NSF DMR 0431350, and LdM 
from the Columbia-Science Po ``Alliance" program. We thank N.~Prokof'ev, A.~Rubtsov and T.~Pruschke for stimulating discussions. The continuous-time calculations were performed on the 
Hreidar cluster at ETH Z\"urich, using the ALPS library \cite{ALPS}.


\begin{thebibliography}{99}
\bibitem{exactdiag} E. Dagotto, Int. J. Mod. Phys. B {\bf 5}, 77 (1991).
\bibitem{auxiliary_problem} E. Y. Loh Jr. \textit{et. al.}, Phys. Rev. B {\bf 41}, 9301 (1990). 
\bibitem{White92}  S. R. White, Phys. Rev. Lett. {\bf 69}, 2863 (1992).
\bibitem{Mueller-Hartmann89} E. M\"uller-Hartmann, Z. Phys. {\bf B74} 507 (1989).
\bibitem{Metzner89} M. Metzner and D. Vollhard, Phys. Rev. Lett. {\bf 62}, 324 (1989).
 \bibitem{Georges92} A. Georges and G. Kotliar Phys. Rev. B 45, 6479 (1992).
\bibitem{Georges96} A. Georges \textit{et al.}, % G. Kotliar, W. Krauth and M. J. Rozenberg, 
Rev. Mod. Phys. {\bf 68}, 13 (1996).
\bibitem{Maier05}  T. Maier \textit{et al.}, %M. Jarrell, T. Pruschke, and M. H. Hettler, 
Rev. Mod. Phys. {\bf 77}, 1027 (2005).
\bibitem{Hirsch86} J. E. Hirsch and R. M. Fye, Phys. Rev. Lett. {\bf 56}, 2521 (1986).  
\bibitem{Caffarel94} M. Caffarel and W. Krauth, Phys. Rev. Lett. {\bf 72},1545 (1994).
\bibitem{Toschi05} A. Toschi \textit{et al.}, % M. Capone, M. Ortolani, P. Calvani, S. Lupi, and C. Castellani, 
Phys. Rev. Lett. {\bf 95}, 097002 (2005). 
%\bibitem{Yoo05} J. Yoo \textit{et al.}, Phys. Rev. B {\bf 71}, 201309(R) (2005). 
\bibitem{Svistunov96} N. V. Prokof'ev \textit{et al.}, % B. V. Svistunov and I. S. Tupitsyn, 
JETP Lett. {\bf 64}, 911 (1996).
\bibitem{Rubtsov05} A. N. Rubtsov \textit{et al.}, %V. V. Savkin and A. I. Lichtenstein, 
Phys. Rev. B {\bf 72}, 035122 (2005).
\bibitem{Werner06b} P. Werner and A. J. Millis, cond-mat/0607136.
\bibitem{Capone06} M. Capone, L. de' Medici, and A. Georges, cond-mat/0512484.
%\bibitem{Feldbacher04} M. Feldbacher \textit{et al.}, Phys. Rev. Lett. {\bf 93} 136405 (2004).
\bibitem{ALPS} M. Troyer {\it et al.}, Lecture Notes in Computer Science {\bf 1505}, 191 (1998); F. Alet \textit{et al.}, J. Phys. Soc. Jpn. Suppl. {\bf 74}, 30 (2005); \url{http://alps.comp-phys.org/} .
\end{thebibliography}
\end{document}